\AtBeginDocument{%
    \newwrite\bibnotes
    \def\bibnotesext{Notes.bib}
    \immediate\openout\bibnotes=\jobname\bibnotesext
    \immediate\write\bibnotes{@CONTROL{REVTEX41Control}}
    \immediate\write\bibnotes{@CONTROL{%
    apsrev41Control,author="08",editor="1",pages="1",title="1",year="1"}}
     \if@filesw
     \immediate\write\@auxout{\string\citation{apsrev41Control}}%
    \fi
}
\documentclass[aps,preprint, superscriptaddress, amsmath, amssymb]{revtex4-1}
\usepackage{float} 
\usepackage{graphicx}
\usepackage{dcolumn}
\usepackage{bm}
\usepackage{upgreek}
\usepackage{times}
\usepackage{mathrsfs}
\usepackage{multirow}
\usepackage{hyperref} 
\usepackage{fancyheadings}
\usepackage[version=3]{mhchem}
\RequirePackage[normalem]{ulem}
\RequirePackage{color}\definecolor{RED}{rgb}{1,0,0}\definecolor{BLUE}{rgb}{0,0,1}

\begin{document}

\title{Stacking defects in GaP nanowires: Electronic structure and optical properties}

\author{Divyanshu Gupta}
\affiliation{Department of Materials Science and Engineering, McMaster University, 1280 Main Street West,
Hamilton, Ontario L8S 4L7, Canada}
\author{Nebile Isik Goktas}
\affiliation{Department of Engineering Physics, McMaster University, 1280 Main Street West,
Hamilton, Ontario L8S 4L7, Canada}
\author{Amit Rao}
\affiliation{Department of Engineering Physics, McMaster University, 1280 Main Street West,
Hamilton, Ontario L8S 4L7, Canada}
\author{Ray LaPierre}
\affiliation{Department of Engineering Physics, McMaster University, 1280 Main Street West,
Hamilton, Ontario L8S 4L7, Canada}
\author{Oleg Rubel}
\email{rubelo@mcmaster.ca}
\affiliation{Department of Materials Science and Engineering, McMaster University, 1280 Main Street West,
Hamilton, Ontario L8S 4L7, Canada}

\date{\today}

\begin{abstract}
Formation of twin boundaries during the growth of semiconductor nanowires is very common. However, the effects of such planar defects on the electronic and optical properties of nanowires are not very well understood. Here, we use a combination of \textit{ab initio} simulation and experimental techniques to study these effects. Twin boundaries in GaP are shown to act as an atomically-narrow plane of wurtzite phase with a type-I homostructure band alignment. Twin boundaries and stacking faults (wider regions of the wurtzite phase) lead to the introduction of shallow trap states observed in photoluminescence studies. These effects should have a profound impact on the efficiency of nanowire-based devices.
\end{abstract}


\maketitle
%
%
\section{Introduction}
III-V semiconductor nanowires (NWs) have applications in electronic, optoelectronic, and photonic devices~\cite{Joyce2011}. III-V NWs can be grown epitaxially on Si making integration of III-V optoelectronic devices with Si-based technology possible~\cite{Zhou2015,Martensson2004,Geum2016}. NWs with embedded quantum dots (e.g. GaAs quantum dots in GaP NWs) have shown potential for use in light-emitting diodes (LEDs), lasers and photodetectors~\cite{Kuyanov2018}.

Crystal imperfections in $\langle111\rangle$   oriented III-V NWs is one of the factors that can limit the performance of optoelectronic devices. Twin boundaries (TBs) are one of the most abundant planar defects observed in NWs~\cite{Johansson2006} as well as bulk semiconductors~\cite{Hurle1995}. A very high twin plane density is usually observed in NWs due to their relatively low stacking fault energy, especially for GaP (30~meV/atom)~\cite{Gottschalk1978}, which can be easily overcome at typical NW synthesis temperatures. Planar crystal defects, such as TBs and stacking faults, can affect electron transport by acting as a carrier scattering source~\cite{Qian2015,Shimamura2013}, a recombination center~\cite{Heiss2011,Vainorius2014,Belabbes2012} or a trap~\cite{Pemasiri2009,Wallentin2012}. Unwanted radiative or non-radiative recombination associated with mid-gap states can be detrimental to the efficiency of optoelectronic devices such as removing carriers from the desired recombination channel in lasers or LEDs or reducing carrier collection in photovoltaic cells. Understanding the potential effects of such defects on the electronic and optical properties of NWs is critical for NW optoelectronic devices.

Here, we present structural and optical studies of GaP NWs combined with \textit{ab initio} calculations to establish a structure-property relationship. Despite GaP being an indirect semiconductor, photoluminescence (PL) spectra of GaP NWs show optical transitions at energies lower than the fundamental band gap of bulk GaP. Transmission electron microscopy (TEM) analysis of NWs indicates that GaP is present in the zinc blende (ZB) phase along with the existence of TBs. We use a density functional theory (DFT) to establish a model of the $\langle111\rangle$  TB in GaP and propose an effective band diagram that explains the origin of sub-band gap optical transitions.
%
%
\section{Method}
\subsection{Experimental details}

\begin{figure}[h]
	\includegraphics{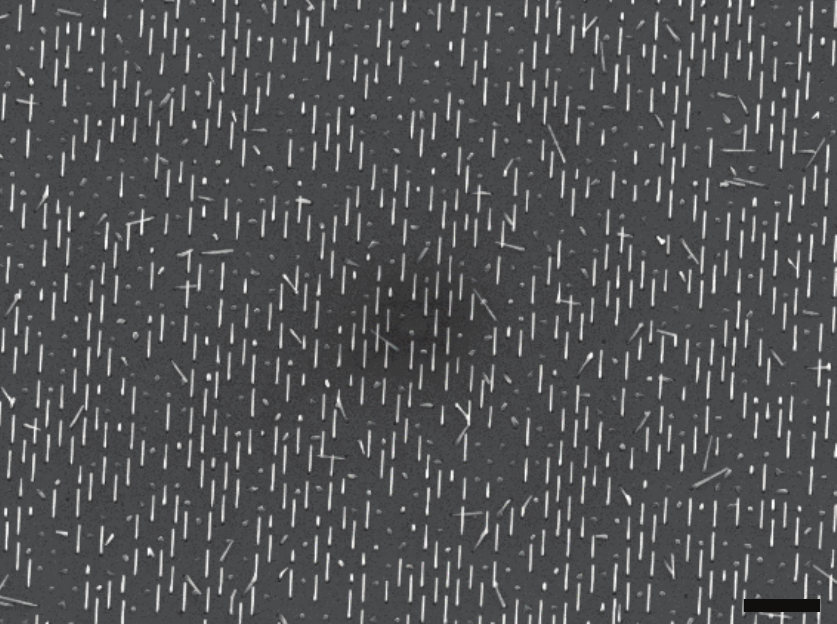}
	\caption{$30^\circ$ tilted view SEM image of GaP NWs. Scale bar is 1~$\mu$m.}
	\label{Fig:1}
\end{figure}

GaP NWs were grown on (111) Si by the self-assisted (seeded by a Ga droplet) selective-area epitaxy method using a multi-source (In,Ga,Al,As,P,Sb) molecular beam epitaxy system. . A solid source effusion cell was used for Ga, while the group V source was $\text{P}_2$ obtained from a high temperature (950$^\circ$C) $\text{PH}_3$ cracker cell. NW growth took place with a Ga impingement rate of 0.125~$\mu$m/h, a V/III flux ratio of 2 and a substrate temperature of 600$^\circ$C. Details of the growth are presented elsewhere~\cite{Kuyanov2017}. After growth, NW arrays were characterized by scanning electron microscopy (SEM) using a JEOL 7000F operating at 5~kV. The SEM image in Fig.~\ref{Fig:1} indicated NWs with $\approx$1~$\mu$m length.

\begin{figure*}[h]
	\includegraphics{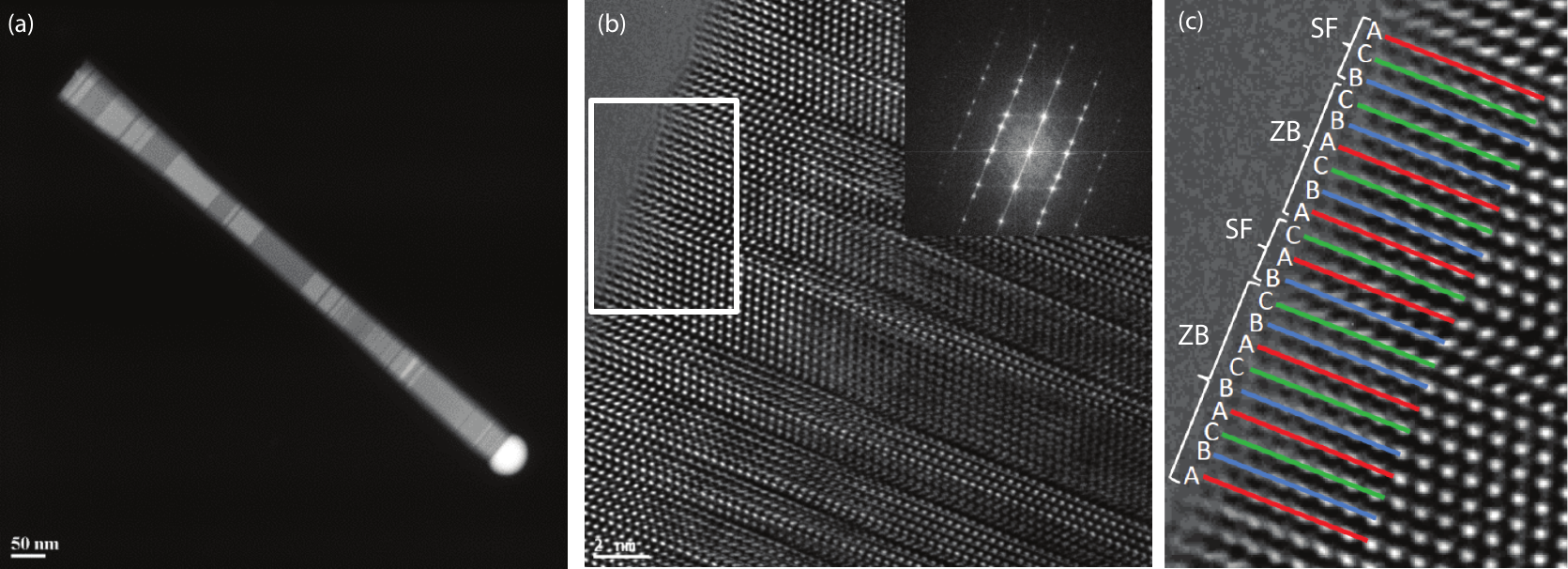}
	\caption{(a) Dark-field STEM image of a single GaP NW. Scale bar is 50 nm. (b) HRTEM image near the sidewall of a single NW. Scale bar is 2 nm. Inset shows the Fourier transform of the image, indicating a twinned ZB crystal structure. (c) Magnified view of the white box in (b), showing the atomic stacking sequence with ZB segments separated by stacking faults (SFs).}
	\label{Fig:2}
\end{figure*}

The NW structure was investigated in a FEI Titan 80-300 LB aberration-corrected scanning transmission electron microscope (STEM) operating at 300 keV. The NWs were removed from their growth substrate for STEM investigation by sonication in a methanol solution followed by transfer to a holey carbon grid. NWs had ~50 nm diameter as evident from the dark-field STEM image in Fig.~\ref{Fig:2}(a). The Ga droplet used for the self-assisted NW growth is evident in the bottom-right of the STEM image. A high-resolution TEM (HRTEM) image in Fig.~\ref{Fig:2}(b) and the Fourier transform of the image (inset) indicate a twinned ZB crystal structure with stacking faults (SFs), which is also evident by the contrast stripes in Fig.~\ref{Fig:2}(a). Figure~\ref{Fig:2}(c) shows the atomic stacking sequence along the $[111]$B growth direction of the NW, showing the ZB twins separated by the SFs.

\begin{figure*}[h]
	\includegraphics{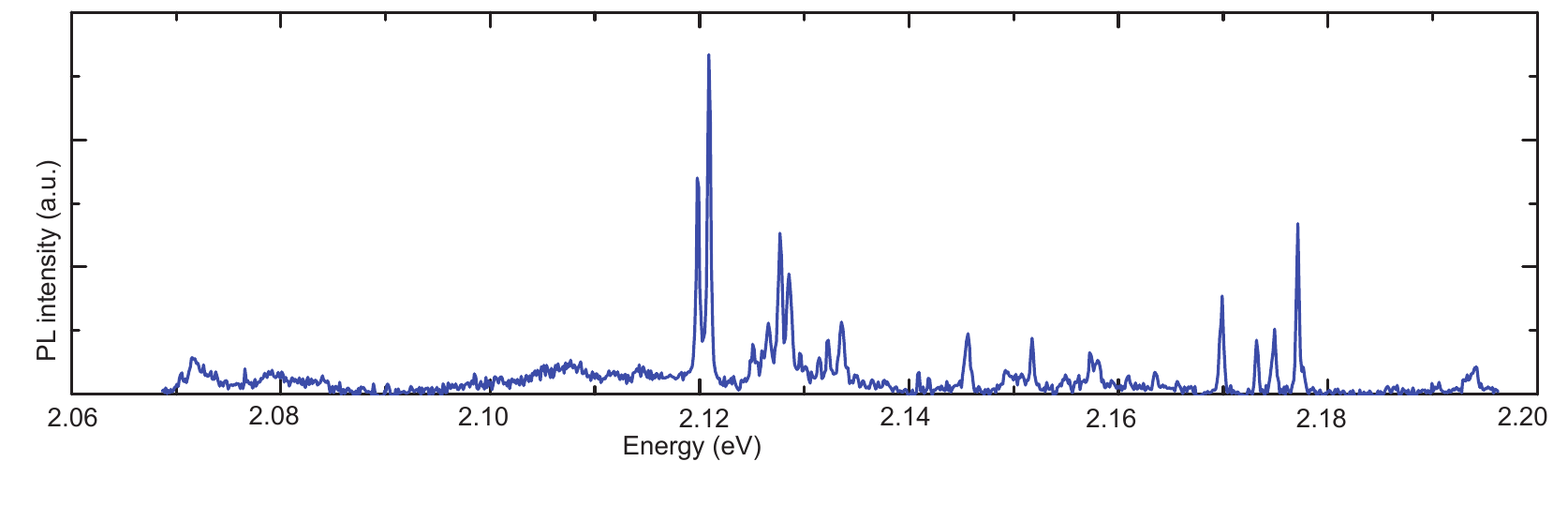}
	\caption{Low temperature (10~K) $\mu$PL spectra obtained from a single GaP NW.}
	\label{Fig:3}
\end{figure*}

Micro-photoluminescence ($\mu$PL) measurements were performed at 10~K on individual NWs. Single NWs were prepared by dispersion on an oxidized Si substrate. An $\text{Ar}^+$ laser with a wavelength of 488~nm was used for excitation through a $60\times$ objective. Spectra were collected by a 0.55~m Horiba Jobin Yvon spectrometer and dispersed onto a $\text{LN}_2$ cooled Si CCD detector. The $\mu$PL spectra (Fig.~\ref{Fig:3}) indicated sharp lines below the band gap of bulk GaP.

\subsection{Computational details}

The first-principles calculations were carried out using DFT~\cite{Kohn1965} and a projector augmented wave method implemented in the Vienna \textit{ab initio} simulation package~\cite{Kresse1999,Bloechl1994,Kresse1996} (VASP). The lattice constant of GaP ZB 2-atom primitive cell was optimized using self-consistent meta-generalized gradient approximation  (SCAN)~\cite{Sun2011} for the exchange and correlation functional. The cut-off energy for the plane-wave expansion was set at 340~eV, which is 25\% higher than the value recommended in pseudopotentials. Ga\_d and P pseudopotentials were used for structural optimizations. The structure was relaxed by minimizing Hellmann-Feynman forces below 2~meV/{\AA}. To capture the localized wurtzite (WZ) region at a twin boundary in a ZB-GaP nanowire, the 4-atom primitive structure of  WZ GaP was lattice matched to ZB along $a$ and $b$ axis while relaxing stress along $c$ axis. The  Brillouin zone of the primitive cells was sampled using $8 \times 8 \times 8$ and $8 \times 8 \times 5$ $k$ meshes for ZB and WZ structures, respectively. The Heyd-Scuseria-Ernzerhof~\cite{Heyd2003}  screened hybrid functional (its HSE-06~\cite{Krukau2006} version) was used in calculations of band gaps and band alignments (VASP tags $\text{HFSCREEN} = 0.2 $ and $\text{AEXX}=0.25$). The Ga\_d potential was substituted by Ga\_vs\_GW potential with $3s$ valence electrons in band alignment calculations when Ga $3s$ reference state was used. The spin-orbit interaction was ignored. The calculated structural parameters and band gaps for ZB and WZ phases of GaP are listed in Table~\ref{Table:1}.

\begin{figure*}[h]
	\includegraphics{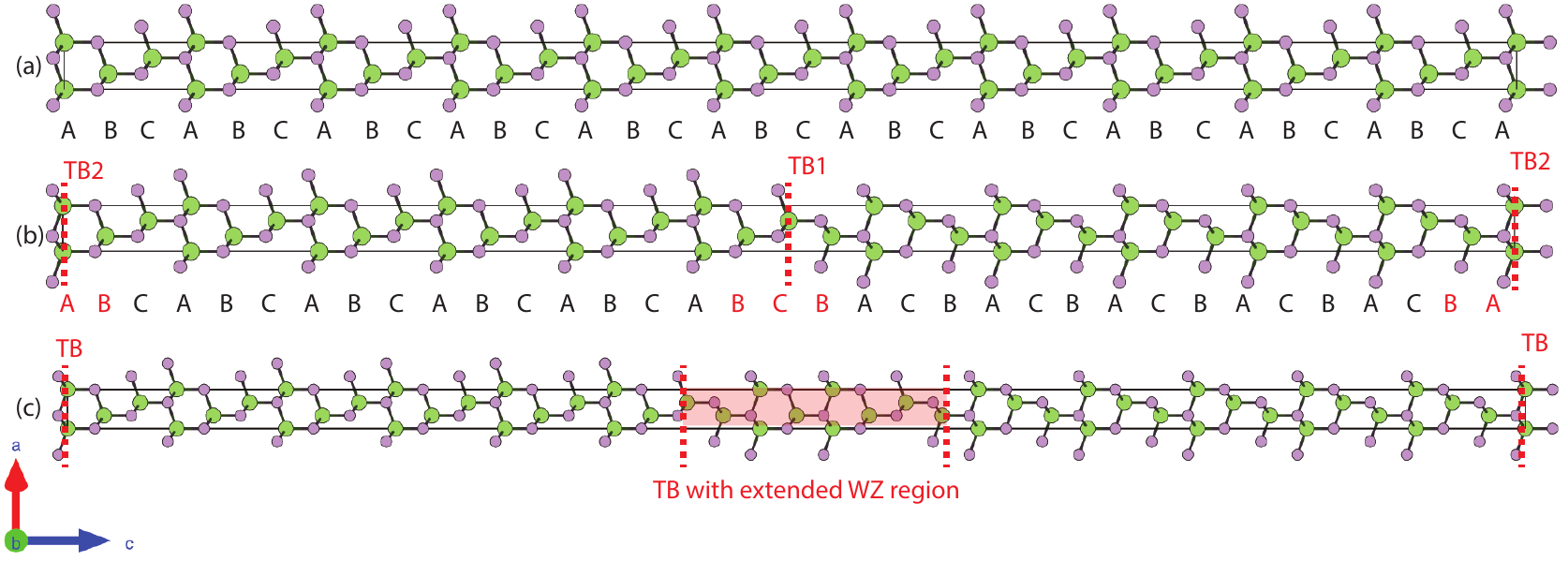}
	\caption{(a) 66-atom GaP-ZB supercell with atomic sequence ABCABC$\ldots$ along [111] direction ($c$ axis) without any defects (CCDC no.~1870794). (b) 68-atom GaP-ZB supercell with 2 twin boundaries TB1 and TB2 (CCDC no.~1870797). (c) 80-atom ZB GaP supercell with stacking fault consisting of an extended WZ region (CCDC no.~1870796).}
	\label{Fig:4}
\end{figure*}

A  supercell of ZB GaP was created along the [111] direction as shown in Fig.~\ref{Fig:4}(a). Along the [111] direction the normal stacking sequence is $\ldots\text{ABCABC}\ldots$. Two 180$^\circ$ twin boundaries were created by altering the stacking sequence as shown in Fig.~\ref{Fig:4}(b). The structures can be accessed at the Cambridge crystallographic data centre (CCDC) under deposition numbers 1870794 and 1870797, respectively. The supercell relaxation was performed by minimizing Hellmann-Feynman forces below 2~meV/{\AA} and stresses below 0.1~kbar using a $6 \times 6 \times 1$ k-mesh grid and the SCAN functional. Subsequently, electronic states were calculated with the HSE-06 functional using a $6 \times 6 \times 1$ k-mesh grid.

To present the band structure of the WZ phase in the f.c.c. Brillouin zone, a supercell was constructed based on a 4-atoms WZ primitive cell by applying the following transformation (rotation) matrix in VESTA software~\cite{Momma2011} 
\begin{equation}\label{Eq:0}
  P =
  \begin{pmatrix}
    1 & 0 & 2 \\
    0 & 1 & 2 \\
    0 & 0 & 3
  \end{pmatrix}.
\end{equation}
An advantage of the resultant 12-atoms supercell is its compatibility with f.c.c. translational lattice vectors (the multiplicity of $1\times1\times6$). The later allows us to use a band structure unfolding~\cite{Popescu2010} as implemented in the fold2Bloch utility~\cite{Rubel2014} and present the WZ band structure in the same Brillouin zone as used for the ZB phase.

%
%
\section{Results and discussion}\label{Sec:Results and discussion}

\begin{figure}[h]
	\includegraphics{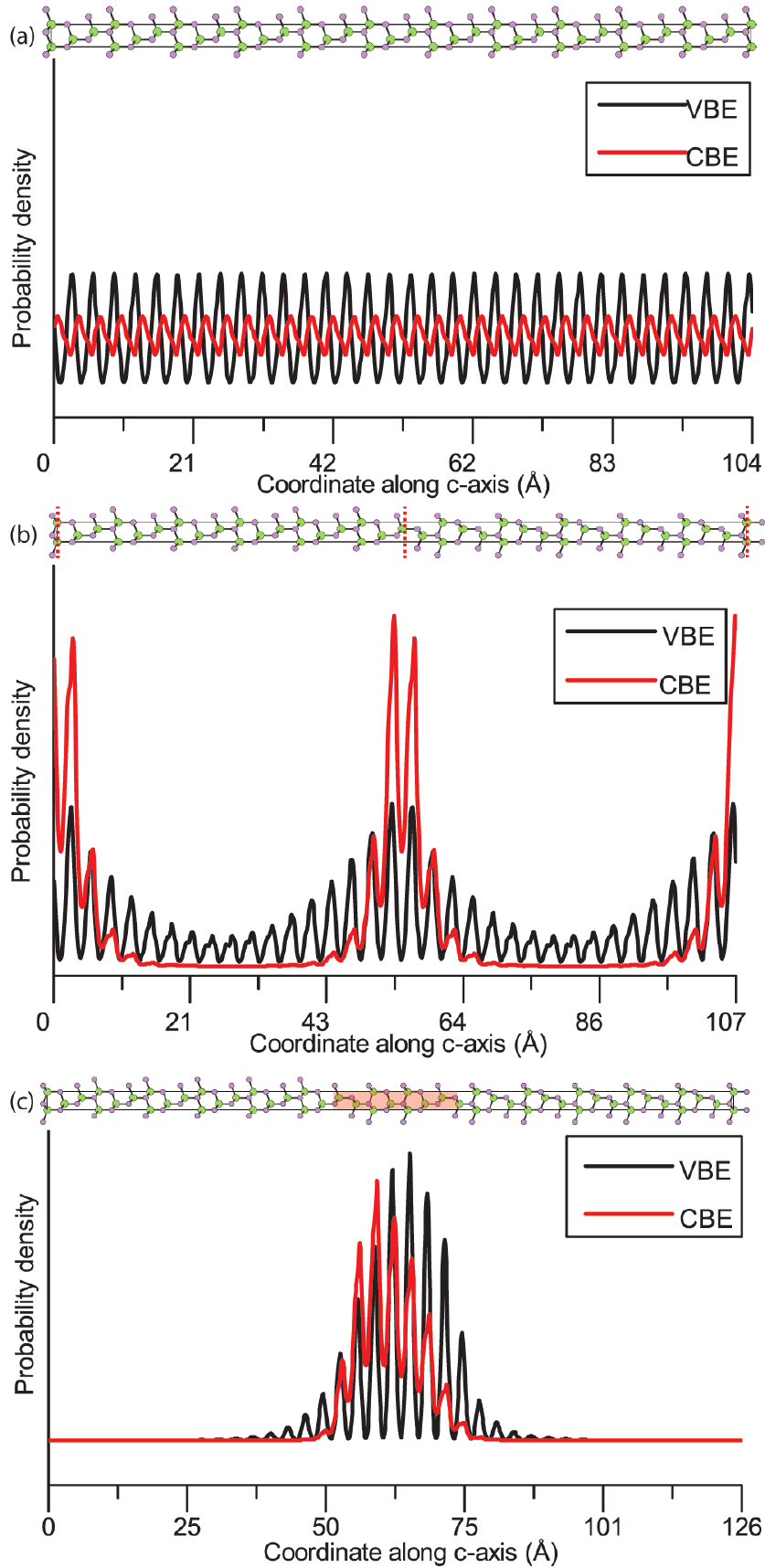}
	\caption{Planar average of DFT-HSE orbitals along $c$ axis for VBE and CBE in (a) defect free structure, (b) structure with twin defects, and (c) structure with extended WZ region.}
	\label{Fig:5}
\end{figure}

Optical transitions with energies lower than the fundamental band gap of 2.34~eV~\cite{Lorenz1968} for bulk GaP (at 10~K) observed in the PL spectrum (Fig.~\ref{Fig:3}) indicate the presence of localized energy states within the band gap, possibly arising due to TBs and SFs. To explore the spatial distribution of charge carriers, the plane averaged wave function amplitude $|\psi(\bm{r})|^2$ distribution in the supercell was calculated using HSE-06~\cite{Krukau2006} exchange and correlation functional.  According to wave function amplitude distributions along the $c$ axis shown in Fig.~\ref{Fig:5}(a), the defect-free structure displays a uniform wave function amplitude distribution in the supercell whereas, for the structure with TBs [Fig.~\ref{Fig:5}(b)], states at the CBE and VBE are both accumulated at the TB. Furthermore, the degree of localization is more prominent for the CBE than for the VBE. Additionally, a structure was constructed to simulate the presence of an extended WZ segment at twin boundaries as illustrated in Fig.~\ref{Fig:5}(c). The wave function amplitude for CBE and VBE is again spatially localized. The degree of localization is similar for CBE and VBE. It is also found that two envelope functions in Fig.~\ref{Fig:5}(c) are asymmetric, indicating the presence of an electric field in the crystal along the $c$ axis. First, we investigate the origin of localization of the electrons and holes at the defect regions.

\begin{figure}[h]
	\includegraphics{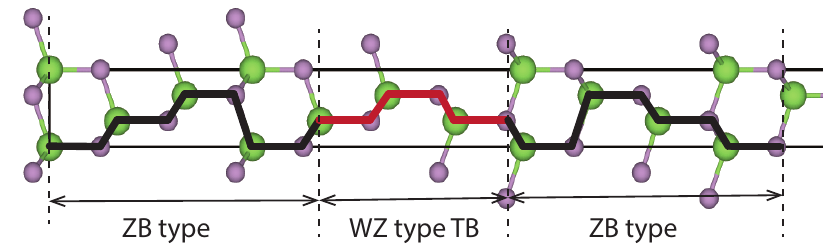}
	\caption{Twin boundary (TB) in ZB resulting in a narrow WZ phase.}
	\label{Fig:6}
\end{figure}

To understand the origin of this localization, the atomic arrangement in the defected supercell was analyzed. Group III-V semiconductors adopt a ZB or WZ type atomic arrangement depending on ionicity of the bond. GaP, being relatively less ionic, naturally exhibits a ZB structure with WZ structure being unstable in bulk. However, in the NW configuration, the WZ structure can become progressively more favourable under certain growth conditions or as the NW diameter decreases~\cite{Joyce2010}. Structures in Fig.~\ref{Fig:4} can be viewed as a sequence of bilayers (one group-III and one group-V atom per bilayer) stacked along the $\langle111\rangle$ direction. The ZB structure follows an $\ldots$ABCABC$\ldots$ stacking sequence of those bilayers as shown in Fig.~\ref{Fig:4}(a), while the WZ phase exhibits an $\ldots$ABAB$\ldots$ sequence where each letter represents a bilayer of the III-V pair. A TB in the ZB phase can be regarded as a stacking fault which creates a local atomically-narrow plane of WZ phase~\citep[p.~55--57]{Li2011} as illustrated in Fig.~\ref{Fig:6}. The co-exsistance of ZB/WZ segments within a structure has been referred to as polytypism. This effect has been studied for multiple systems such as GaAs\cite{Ahtapodov2016},InP\cite{Wang2016}, InAs\cite{Caroff2008}, ZnO\cite{Aebersold2017}, Si\cite{Liu2015} and other III-V and II-VI compounds and has been shown to have a significant impact on the optical and electronic properties of the materials. 

Electronic and optical properties of the WZ phase are very different from its ZB counterpart. This difference could explain the unexpected PL transitions. Previous experimental and theoretical investigations suggest that the GaP WZ structure is a pseudo-direct band gap material with a lower band gap of 2.13~eV (at 4~K) as compared to 2.34~eV (at 10~K) for the GaP ZB structure~\cite{Belabbes2019}~\cite{Assali2013}. To predict the effects of narrow WZ regions in ZB phase, it is crucial to establish the relative alignment of CBE and VBE in the two phases.

Prior theoretical studies suggest that III-V semiconductors in a WZ/ZB homostructure generally follow a type-II band alignment~\cite{Murayama1994}. This type of the band alignment will favour a spatial separation of electrons and holes at the WZ/ZB interface. In the case of GaP, literature shows a disagreement about the band alignment between the ZB and WZ phase. An early theoretical work by \citet{Murayama1994} proposed a type-II alignment, whereas later results by \citet{Belabbes2012} claim a type-I alignment. Polarization-dependent PL measurement reported by \citet{Assali2017} support a type-I alignment. It should be noted that \citet{Murayama1994} used a local density approximation for the exchange and correlation functional for calculation of band offsets leading to a significant underestimation of band gaps for ZB as well as WZ phases. Furthermore, the WZ structures used for these calculations were completely relaxed, which would not be applicable to a homojunction because the WZ present at the TB would be lattice matched to ZB phase along $a$ and $b$ axis leading to an epitaxial strain in the $a$-$b$ plane, while the relaxation of stress along the $c$ axis is permitted. This necessitates a further investigation into the band alignment pertaining to the present case.

\begin{figure}[h]
	\includegraphics{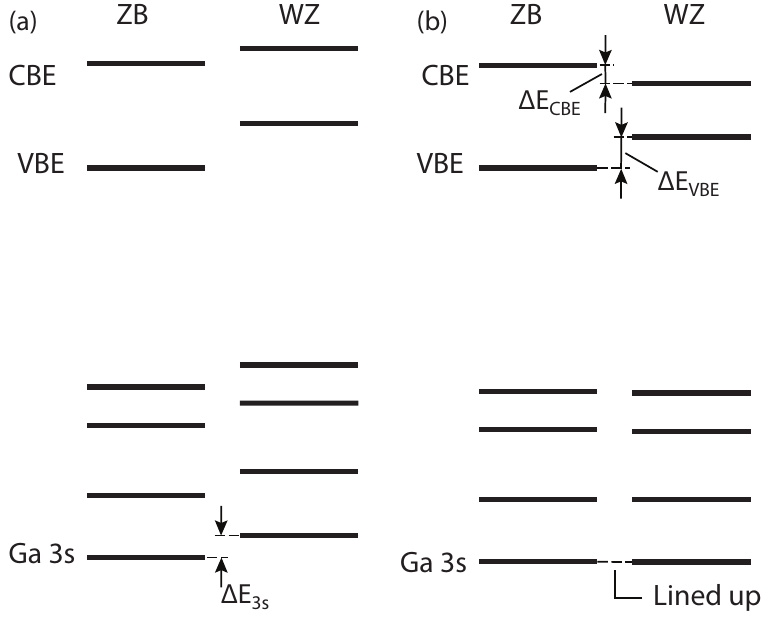}
	\caption{Relative arrangement of energy levels in ZB and WZ GaP primitive unit cell (a) prior to band alignment and (b) after the alignment of Ga $3s$ states.}
	\label{Fig:7}
\end{figure}

An average Coulomb potential serves as a reference energy in DFT calculations of solids with periodic boundary conditions. This reference energy is ill-defined due to the lack of vacuum. Thus, band alignments cannot be derived directly from eigenvalues of ZB and WZ structures obtained from separate calculations. The first step to reliably determine the relative position of band edges in the two phases is to identify a reference state that does not change with the crystal structure. Core states are an ideal candidate for this purpose as these states are least sensitive to local coordination of atoms and thus should be virtually unaffected by changes in the valence electron density (see Appendix for a rigorous evaluation of the associated inaccuracy). This approach was successfully applied to the calculation of volume deformation potentials in III-V and II-VI semiconductors~\cite{Wei1999}. Eigenvalues of the $3s$ state of Ga are obtained for bulk ZB and WZ structures from separate calculations (WZ is lattice matched to ZB along $a$ and $b$-axis, and allowed to relax along $c$ axis). Figure~\ref{Fig:7}(a) shows a qualitative representation of the alignment of eigenvalues in ZB and WZ structures. Note that Ga $3s$ states are slightly misaligned due to uncertainties in the reference energy between the two calculations. The difference in $3s$ state of Ga is adjusted by shifting the WZ bands by
\begin{equation}
\Delta E_{3s}= E_{3s}^\text{Ga,WZ} - E_{3s}^\text{Ga,ZB}.
\label{Eq:1}	
\end{equation}
The adjusted band alignment is displayed in Fig.~\ref{Fig:7}(b). The discontinuity in the valence band ($\Delta E_\text{VBE}$) and that in the conduction band ($\Delta E_\text{CBE}$) are evaluated as:
\begin{equation}
\Delta E_\text{VBE/CBE} \approx E_\text{VBE/CBE}^\text{WZ} - E_\text{VBE/CBE}^\text{ZB} - \Delta E_{3s}.
\label{Eq:2}	
\end{equation}
Calculations using HSE-06 exchange-correlation functional yield $\Delta E_\text{VBE} = 137$~meV and $\Delta E_\text{CBE} = -20$~meV. In comparison, \citet{Belabbes2012} obtained  $\Delta E_\text{VBE} = 135$~meV and $\Delta E_\text{CBE} = -14$~meV with a completely relaxed WZ structure. The approximate sign in Eq.~(\ref{Eq:2}) indicates a small inaccuracy of $\sim5$~meV in the $\Delta E_{3s}$ term associated with a shift of $E_{3s}^\text{Ga}$ energies at the WZ/ZB interface (see Appendix for details).

\begin{figure}[h]
	\includegraphics{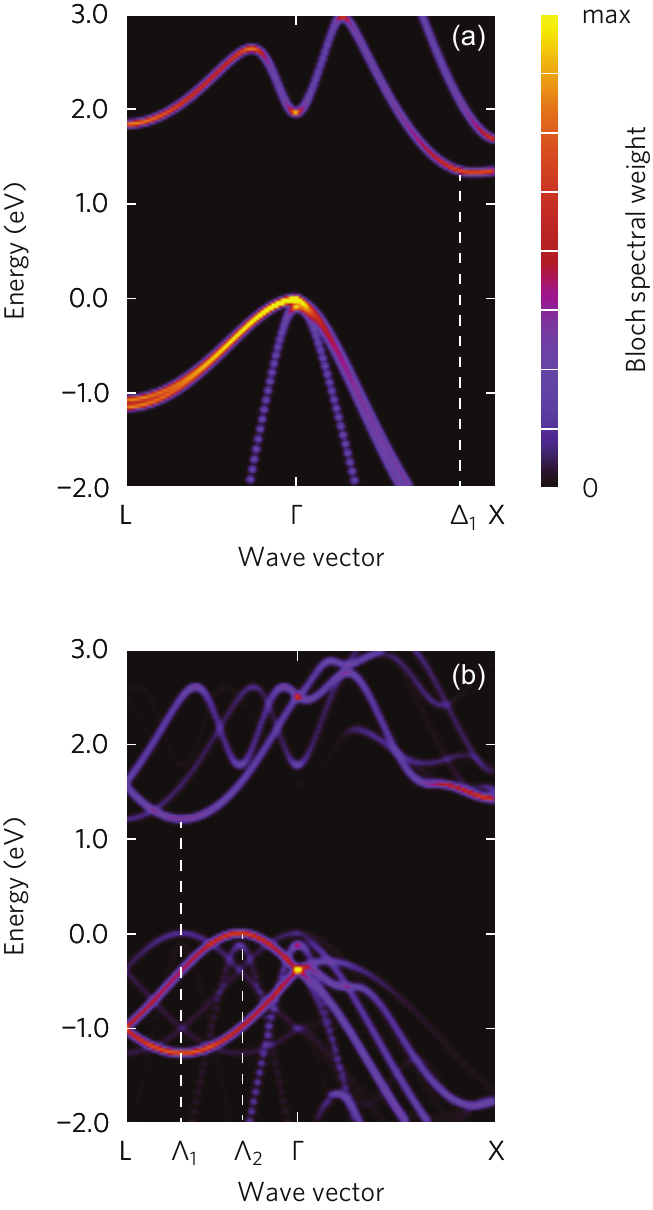}
	\caption{Band structure of GaP in (a) ZB phase and (b) WZ phase presented in the f.c.c. Brillouin zone. Band edges of the two structures are at different $k$ points. The WZ phase also shows an indirect band gap, which explains why its lowest energy transition is dipole-forbidden~\cite{Belabbes2019} even though the band structure appears direct when presented in the hexagonal Brillouin zone (both band edges fall into the $\Gamma$ point). Calculations were performed with the PBEsol~\cite{Perdew2008} exchange correlation functional. An origin of the energy scale is set at the Fermi energy.}
	\label{Fig:8}
\end{figure}

Calculations with HSE-06 functional indicate a type-I band alignment for the WZ/ZB GaP homojunction which agrees with the spatial localization of CBE and VBE states in Fig.~\ref{Fig:5}. However, a stronger spatial localization of the CBE state in Fig.~\ref{Fig:5} suggests their stronger quantum confinement, which seems contradictory to the band alignment results. It should be noted that the alignment calculations are done for the band edges without paying attention to the band character. In ordinary heterostructures, \textit{e.g.}, (InGa)As/GaAs, band edges of the quantum well and the barrier material are both located in the same $k$ point of the Brillouin zone. However, this is not the case in the WZ/ZB GaP homostructure where the band edges belong to different points in the f.c.c. Brillouin zone ($\Delta_1$ \textit{vs} $\Lambda_1$ for the CBE) as shown in Fig.~\ref{Fig:8}. Alternatively, the band alignment in the conduction band can be evaluated at the same $\Lambda_1$ point for both structures, resulting in a much deeper confinement potential for electrons. A similar situation takes place in GaAs/AlAs quantum wells, where it is customary to consider an electron localized in a GaAs domain as confined by a potential barrier corresponding to the energy difference between $\Gamma_{1\text{c}}$ states of AlAs and GaAs~\cite{Franceschetti1995}.

The accuracy of band alignment is further validated by using a more traditional and well established method laid down by \citet{VandeWalle1987}, which involves examining the macroscopic average of the electrostatic potential determined in the bulk-like regions of each phase in a supercell (further details can be found in the Appendix). The two methods showed values of the band discontinuity within 10~meV of one another.

\begin{figure}[h]
	\includegraphics{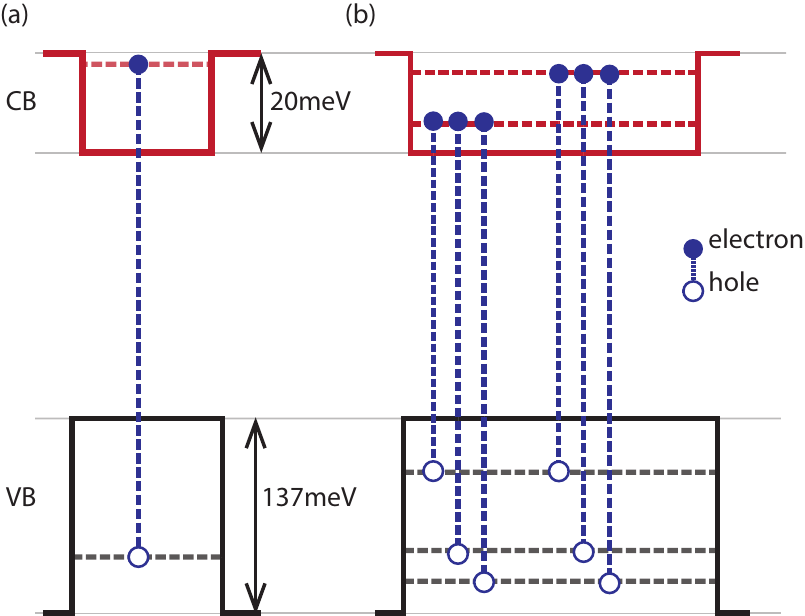}
	\caption{Quantization of energy levels in (a) narrow and (b) wide WZ regions embedded in ZB phase.}
	\label{Fig:9}
\end{figure}

\begin{figure*}[h]
	\includegraphics{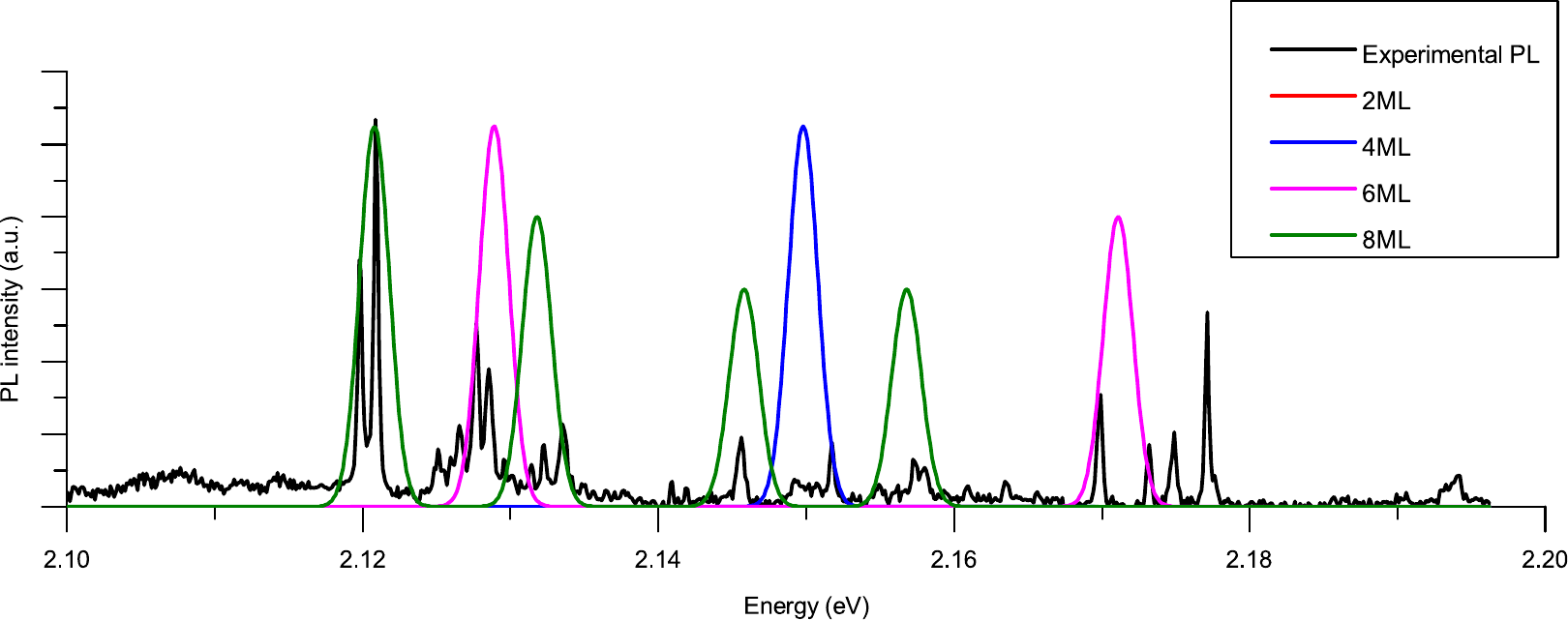}
	\caption{Comparison of experimental $\mu$PL spectrum and calculated electronic transition in WZ/ZB GaP quantum wells of variable width [Fig.~\ref{Fig:9}(b)].}
	\label{Fig:10}
\end{figure*}

The calculated band offsets can now be used to assess a quantization energy. In GaP-WZ/ZB homostructures, the trap states are relatively shallow [Fig.~\ref{Fig:9}(a)] due to the narrow width of TBs and the associated quantization of the electron and hole states. The energy of recombination of excitations trapped at TBs can be calculated using a solution of the Schr{\"o}dinger wave equation for a one-dimensional finite potential well as follows~\cite{Aronstein2000}:
\begin{equation}
\tan \left(\frac{L\sqrt{2mE}}{2\hbar} \right) = \sqrt{\frac{V-E}{E}}
\label{Eq:3}	
\end{equation}
where $\hbar$ is the reduced Plank's constant, $m$ is the effective mass of hole (electron), $V$ is the depth of potential well, \textit{i.e.} $\Delta E_\text{VBE}$ or $\Delta E_\text{CBE}$, $E$ is the quantized energy level within the well and $L$ is the width of the well. The effective mass of electrons is taken as 1.12$m_{e}$~\citep[p.~104]{Sh1996} and the effective mass of holes (heavy hole) is taken as 0.79$m_{e}$~\citep[p.~104]{Sh1996}, where $m_{e}$ is the rest mass of the electron. The recombination energies are obtained for varying potential well widths starting at 2~WZ monolayers (MLs) ($L=6.32$~{\AA}), \textit{i.e.}, a twin boundary, and widening with 2~ML intervals to simulate regions with extended WZ segments. Transition energies were calculated to a width of 8~MLs beyond which, the difference in energy levels was less than 1~meV.  Qualitatively, we see a higher number of states within the well  along with progressively lower recombination energies as the potential well widens [Fig.~\ref{Fig:9}(b)]. The transition energies thus obtained are plotted against the  experimental $\mu$PL spectrum in Fig.~\ref{Fig:10}. A Gaussian broadening with a full width at half maximum of 1~meV has been assigned to the calculated transition energies. The calculated spectrum was red-shifted by 35~meV to account for a minor inaccuracy in the band gap calculated by HSE-06 and possible excitonic effects. The transition arising from an isolated twin boundary is expected near 2.22~eV. We consider only heavy holes for our calculation as transition energies arising from light holes (effective mass 0.14$m_{e}$~\citep[p.~104]{Sh1996}) were found to be higher than the range of experimental data. The transition energies calculated using a simple finite potential well model are in agreement with the experimental data (Fig.~\ref{Fig:10}).

Finally, we discuss the asymmetric nature of the probability density distribution seen in Fig.~\ref{Fig:5}(c). Group III-V WZ structures exhibit a spontaneous polarization along (0001) axis~\cite{Bernardini1997}, resulting in a built-in electric field in ZB and WZ segments that causes a charge segregation at the WZ/ZB interface. This effect has been widely studied~\cite{Bernardini1997,Wagner2002,Park2018} in various III-V compounds. It has also been shown that change in absolute as well as relative thickness of ZB and WZ phases lead to variation in the intensity of the built-in electric field. Wider WZ segments result in a stronger electric field and, hence, more prominent charge segregation~\cite{Zhang2017}. Similar behaviour is expected for GaP. This segregation is expected to be present at a twin boundary as well as adjacent SFs acting as wide WZ regions with a higher electric field as compared to an isolated twin boundary, which leads to a stronger segregation of charge carriers seen in Fig.~\ref{Fig:5}(c). 
%
%
\section{Conclusion}
The effect of twin boundaries and stacking faults on electrical and optical properties of GaP nanowires was studied using experimental techniques in combination with \textit{ab initio} density functional simulations. Transmission electron microscopy examination of GaP nanowires shows the presence of a zinc blende phase with twin boundaries and extended WZ segments. Photoluminescence studies of the nanowires show the presence of radiative recombination below the fundamental band gap of zinc blende GaP. Twin boundaries can be viewed as an atomically narrow wurtzite phase. \textit{Ab initio} calculations suggest that the WZ/ZB GaP homostructure shows a type-I band alignment. Thus, the twin boundary acts as shallow trap for electrons and holes. The energy of recombination of excitations trapped at twin boundaries is expected to be approximately 60~meV below the fundamental band gap of zinc blende GaP. Photoluminescence lines with lower energies can arise from stacking faults (a few monolayers of extended wurtzite region) at the twin boundaries.

\appendix*

\section{Validation of band alignment calculations}

\begin{figure}[h]
	\includegraphics{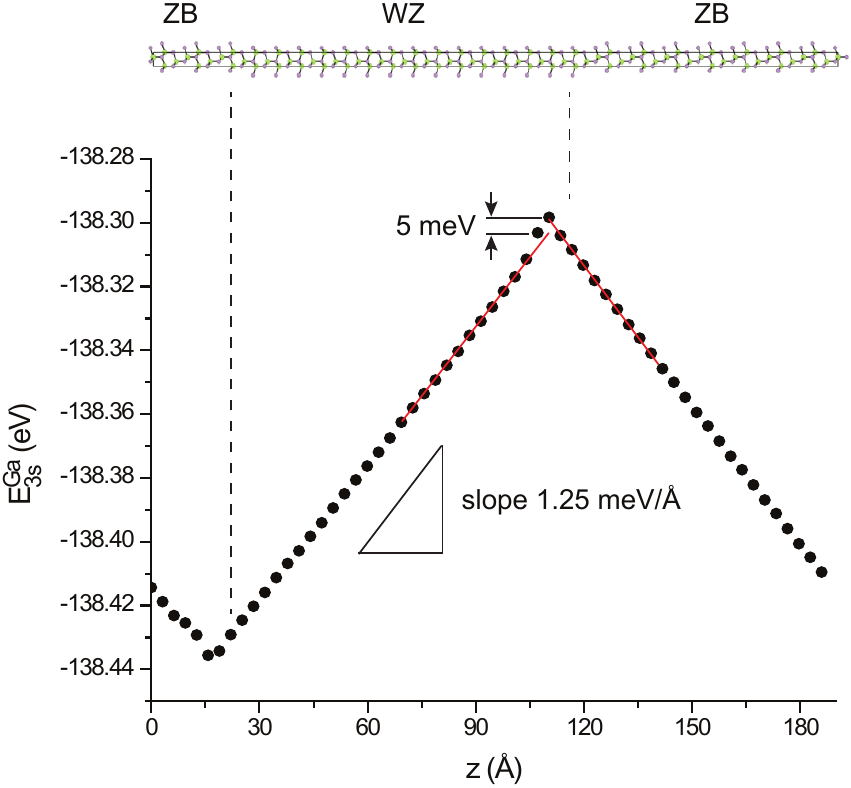}
	\caption{Variation of Ga $3s$ core energy levels along $c$ axis in a 1:1 ZB:WZ GaP homostructure (top panel) composed of 120 atoms.}
	\label{Fig:11}
\end{figure}

\textit{Average electrostatic potential}: Band alignment determination using an averaged electrostatic potential was carried out according to the procedure outlined by \citet{Weston2018} as an alternative to the alignment of Ga $3s$ core states used in the main part of the paper. Calculations were performed with VASP package employing PBEsol~\cite{Perdew2008} exchange and correlation functional. First, the position of the valence band maxima is calculated with respect to the average electrostatic potential $\langle V\rangle^\text{ZB(WZ)}_\text{B}$ for both ZB and WZ structures in bulk. Further, the average electrostatic potential $\langle V\rangle^\text{ZB(WZ)}_\text{H}$ is determined for bulk-like ZB and WZ regions within a 1:1 ZB:WZ homostructure supercell with 120 atoms (not relaxed) shown in the upper panel of Fig.~\ref{Fig:11}. The relative misalignment of the electrostatic potential is expressed as:
\begin{equation}
  \Delta V=\langle V\rangle^\text{WZ}_\text{B} - \langle V\rangle^\text{ZB}_\text{B} - \langle V\rangle^\text{WZ}_\text{H} + \langle V\rangle^\text{ZB}_\text{H}.
\end{equation}
Finally, the valence band offset $\Delta E_\text{VBE}$ is defined as:
\begin{equation}
 \Delta E_\text{VBE}  = E_\text{VBE}^\text{WZ} - E_\text{VBE}^\text{ZB} - \Delta V.
\end{equation}
The calculated band offset $\Delta E_\text{VBE}=110$~meV agrees with the corresponding value of 118~meV determined with PBEsol using Ga $3s$ core states as the reference [Eq.~(\ref{Eq:2})]. Since, both methods yield the same results, we prefer using the core states as a reference due to its simplicity.

\textit{Misalignment of core states}: When Ga $3s$ core states are used as a reference for determining the band alignment, it is assumed that the $E^\text{Ga}_{3s}$ energy in GaP is not sensitive to the crystal structure (ZB \textit{vs} WZ). Accuracy of this approximation can be further tested by analyzing a discontinuity of $E^\text{Ga}_{3s}$ at the WZ/ZB interface of the 120 atoms supercell discussed in the previous paragraph. The calculations were performed with the WIEN2k~\cite{Blaha2018}, an all electron density functional package. The WIEN2k basis set is better suited than VASP for this purpose, since core states are confined within muffin tin spheres centred at individual atoms allowing for an easy link between $E^\text{Ga}_{3s}$ values and atomic coordinates. The radii ($R^\text{MT}$) of the muffin tin spheres are chosen to be equal to 2.36  and 1.93~Bohr for Ga and P, respectively. The product of the minimum radius and the maximum cut-off wave vector in the reciprocal space was set at the value of $R^\text{MT}_{\text{min}}K_{\text{max}} = 7$. The Brillouin zone was sampled using an $8 \times 8 \times 1$ $k$-point mesh. Evolution of $E^\text{Ga}_{3s}$ core energy levels along $c$ axis of the supercell is shown in Fig.~\ref{Fig:11}. A discontinuity in the $E^\text{Ga}_{3s}$ energy level at the ZB/WZ interface is about 5~meV, which sets the error bar for band offsets obtained with this method.

The slope of energy levels in Fig.~\ref{Fig:11} indicates presence of an electric field of equal magnitude 1.25~meV/{\AA}, but opposite direction, in both WZ and ZB regions of the model. The field originates likely due to a spontaneous polarization present in the WZ structure. The effect becomes less notable in structures with narrow WZ regions as discussed at the end of Sec.~\ref{Sec:Results and discussion}.

%
%
\begin{acknowledgments}
Authors would like to acknowledge funding provided by the Natural Sciences and Engineering Research Council of Canada under the Discovery Grant Programs RGPIN-2015-04518 (D.G. and O.R.) and RGPIN-2018-04015 (N.G. and R.L.). DFT calculations were performed using a Compute Canada infrastructure supported by the Canada Foundation for Innovation under John R. Evans Leaders Fund.
\end{acknowledgments}

%
%

%

%
%
\newpage

\begin{table}
  \caption{Calculated and experimental structural parameters ($a$ and $c$) and band gap ($E_\text{g}$) of GaP in two phases.}\label{Table:1}
  \begin{ruledtabular}
    \begin{tabular}{llcccc}
      Phase & Parameter & Theory  &  Experimental \\
       &  & (VASP)  &  (at $T$) \\
      \hline
       ZB & $a$ ({\AA}) &  $5.449$ & 5.446 (300~K)~\cite{Treece1992} \\
        & $E_\text{g}$ (eV) &  2.30 & 2.34 (10~K)~\cite{Lorenz1968} \\
      \hline
       WZ on ZB & $a$ ({\AA})  & 3.853\footnote{The value is set at $a^\text{WZ}=a^\text{ZB}/\sqrt{2}$ to maintain the interface with ZB structure while $c^\text{WZ}$ is allowed to relax (CCDC no.~1870795).} & $-$ \\
        & $c$ ({\AA}) & 6.339 & $-$  \\
        & $E_\text{g}$ (eV) & 2.15  & 2.13 (4~K)~\cite{Halder2018} \\
    \end{tabular}
  \end{ruledtabular}
\end{table}

%
%
\clearpage

\end{document}